\pdfoutput=1
\documentclass{JINST}

\title{Measurement of antiproton annihilation on Cu, Ag and Au with emulsion films}

\author{
S. Aghion$^{a,b}$,
C. Amsler$^{c,d}$,
A. Ariga$^{c}$,
T. Ariga$^c$\thanks{Corresponding author.}~, 
G. Bonomi$^{e,f}$,
P. Br\"aunig$^{g}$,
R. S. Brusa$^{h,i}$,
L. Cabaret$^{j}$,
M. Caccia$^{b,k}$,
R. Caravita$^{l,m}$,
F. Castelli$^{b,n}$,
G. Cerchiari$^{o}$,
D. Comparat$^{j}$,
G. Consolati$^{a,b}$,
A. Demetrio$^{g}$,
L. Di Noto$^{l,m}$,
M. Doser$^{p}$,
A. Ereditato$^{c}$,
C. Evans$^{a,b}$,
R. Ferragut$^{a,b}$,
J. Fesel$^{p}$,
A. Fontana$^{f}$,
S. Gerber$^{p}$,
M. Giammarchi$^{i}$,
A. Gligorova$^{q}$,
F. Guatieri$^{h,i}$,
S. Haider$^{p}$,
A. Hinterberger$^{p}$,
H. Holmestad$^{r}$,
T. Huse$^{r}$,
J. Kawada$^{c}$,
A. Kellerbauer$^{o}$,
M. Kimura$^{c}$,
D. Krasnick\'y$^{l,m}$,
V. Lagomarsino$^{l,m}$,
P. Lansonneur$^{s}$,
P. Lebrun$^{s}$,
C. Malbrunot$^{d,p}$,
S. Mariazzi$^{d}$,
V. Matveev$^{t,u}$,
Z. Mazzotta$^{b,n}$,
S. R. M\"{u}ller$^{g}$,
G. Nebbia$^{v}$,
P. Nedelec$^{s}$,
M. Oberthaler$^{g}$,
N. Pacifico$^{q}$,
D. Pagano$^{e,f}$,
L. Penasa$^{h,i}$,
V. Petracek$^{w}$,
C. Pistillo$^{c}$,
F. Prelz$^{b}$,
M. Prevedelli$^{x}$,
L. Ravelli$^{h,i}$,
B. Rienaecker$^{p}$,
O. M. R{\o}hne$^{r}$,
A. Rotondi$^{f,y}$,
M. Sacerdoti$^{b,n}$,
H. Sandaker$^{r}$,
R. Santoro$^{b,k}$,
P. Scampoli$^{c,z}$,
M. Simon$^{d}$,
L. Smestad$^{p,aa}$,
F. Sorrentino$^{l,m}$,
G. Testera$^{m}$,
I. C. Tietje$^{p}$,
S. Vamosi$^{d}$,
M. Vladymyrov$^{c}$,
E. Widmann$^{d}$,
P. Yzombard$^{j}$,
C. Zimmer$^{o,p,bb}$,
J. Zmeskal$^{d}$,
N. Zurlo$^{f,cc}$
\\
\llap{$^a$}{Politecnico of Milano, Piazza Leonardo da Vinci 32, 20133 Milano, Italy}\\
\llap{$^b$}{INFN Milano, via Celoria 16, 20133, Milano, Italy}\\
\llap{$^c$}{Laboratory for High Energy Physics, Albert Einstein Center for Fundamental Physics, University of Bern, 3012 Bern, Switzerland}\\
\llap{$^d$}{Stefan Meyer Institute for Subatomic Physics, Austrian Academy of Sciences, Boltzmanngasse 3, 1090 Vienna, Austria}\\
\llap{$^e$}{Department of Mechanical and Industrial Engineering, University of Brescia, via Branze 38, 25123 Brescia, Italy}\\
\llap{$^f$}{INFN Pavia, via Bassi 6, 27100 Pavia, Italy}\\
\llap{$^g$}{Kirchhoff-Institute for Physics, Heidelberg University, Im Neuenheimer Feld 227, 69120 Heidelberg, Germany}\\
\llap{$^h$}{Department of Physics, University of Trento, via Sommarive 14, 38123 Povo, Trento, Italy}\\
\llap{$^i$}{TIFPA/INFN Trento, via Sommarive 14, 38123 Povo, Trento, Italy}\\
\llap{$^j$}{Laboratoire Aim\'e Cotton, Universit\'e Paris-Sud, ENS Cachan, CNRS, Universit\'e Paris-Saclay, 91405 Orsay Cedex, France}\\
\llap{$^k$}{Department of Science, University of Insubria, Via Valleggio 11, 22100 Como, Italy}\\
\llap{$^l$}{Department of Physics, University of Genova, via Dodecaneso 33, 16146 Genova, Italy}\\
\llap{$^m$}{INFN Genova, via Dodecaneso 33, 16146 Genova, Italy}\\
\llap{$^n$}{Department of Physics, University of Milano, via Celoria 16, 20133 Milano, Italy}\\
\llap{$^o$}{Max Planck Institute for Nuclear Physics, Saupfercheckweg 1, 69117 Heidelberg, Germany}\\
\llap{$^p$}{Physics Department, CERN, 1211 Geneva 23, Switzerland}\\
\llap{$^q$}{Institute of Physics and Technology, University of Bergen, All\'egaten 55, 5007 Bergen, Norway}\\
\llap{$^r$}{Department of Physics, University of Oslo, Sem S{\ae}lands vei 24, 0371 Oslo, Norway}\\
\llap{$^s$}{Institute of Nuclear Physics, CNRS/IN2p3, University of Lyon 1, 69622 Villeurbanne, France}\\
\llap{$^t$}{Institute for Nuclear Research of the Russian Academy of Science, Moscow 117312, Russia}\\
\llap{$^u$}{Joint Institute for Nuclear Research, 141980 Dubna, Russia}\\
\llap{$^v$}{INFN Padova, via Marzolo 8, 35131 Padova, Italy}\\
\llap{$^w$}{Czech Technical University, Prague, Brehov$\acute{a}$ 7, 11519 Prague 1, Czech Republic}\\
\llap{$^x$}{University of Bologna, Viale Berti Pichat 6/2, 40126 Bologna, Italy}\\
\llap{$^y$}{Department of Physics, University of Pavia, via Bassi 6, 27100 Pavia, Italy}\\
\llap{$^z$}{Department of Physics ``Ettore Pancini'', University of Napoli Federico II, Complesso Universitario di Monte S. Angelo, 80126, Napoli, Italy}\\
\llap{$^{aa}$}{The Research Council of Norway, P.O. Box 564, NO-1327 Lysaker, Norway}\\
\llap{$^{bb}$}{Department of Physics and Astronomy, Heidelberg University, Im Neuenheimer Feld 227, 69120 Heidelberg, Germany}\\
\llap{$^{cc}$}{Department of Civil Engineering, University of Brescia, via Branze 43, 25123 Brescia, Italy}\\

E-mail: \email{tomoko.ariga@lhep.unibe.ch}}

\abstract{The characteristics of low energy antiproton annihilations on nuclei (e.g. hadronization and product multiplicities) are not well known, and Monte Carlo simulation packages that use different models provide different descriptions of the annihilation events. In this study, we measured the particle multiplicities resulting from antiproton annihilations on nuclei. The results were compared with predictions obtained using different models in the simulation tools GEANT4 and FLUKA. For this study, we exposed thin targets (Cu, Ag and Au) to a very low energy antiproton beam from CERN's Antiproton Decelerator, exploiting the secondary beamline available in the AEgIS experimental zone. The antiproton annihilation products were detected using emulsion films developed at the Laboratory of High Energy Physics in Bern, where they were analysed at the automatic microscope facility. The fragment multiplicity measured in this study is in good agreement with results obtained with FLUKA simulations for both minimally and heavily ionizing particles.}


\begin{document}

\section{Introduction}
\label{sec:intro}

Emulsion films have recently been considered as possible position detectors for low-energy antimatter studies. These studies include the AEgIS (AD6) experiment at CERN~\cite{aegisprop,AEgIS_emul1,AEgIS_emul2,nature}, whose goal is the measurement of the Earth's gravitational acceleration on antihydrogen atoms. Another collaboration proposed emulsions for their studies on positrons, as described in~\cite{quplas}. In particular, in the case of the AEgIS experiment, the position-sensitive detector must have a micrometer-level resolution to allow the required sensitivity of $\sim$1\% for the gravitational acceleration measurement. 
Spatial resolutions of $\sim$1-2~$\mu$m can be achieved with emulsion films~\cite{ereditato}, and they have been exploited before for the reconstruction of antihydrogen impact points from annihilation products~\cite{AEgIS_emul2}. Films with this resolution, combined with a time of flight detector, could allow the experimental goal to be achieved. In the same paper, a preliminary study of antiproton-nuclei annihilations was also reported. 
That study assessed particle multiplicities resulting from antiproton annihilations on emulsion films and aluminium. Recently, again within the context of the AEgIS experiment, similar measurements were performed by means of a silicon detector also acting as an annihilation target~\cite{silicon}.
Apart from the obvious applications in nuclear physics, measuring the decay products of low-energy antiproton annihilation in different materials provides a useful check of the ability of standard Monte Carlo packages to reproduce fragment multiplicities, type and energy distributions stemming from antiproton (or antineutron) annihilations on nuclei at rest. Although measurements of the multiplicities of pions and other charged particles with energies higher than $\sim$ 50 MeV are available in the literature~\cite{ref1994,ref1995}, the production of highly ionizing nuclear fragments with short range has not been studied sufficiently. A measurement of the multiplicities of charged products in antiproton-aluminium annihilations was reported in~\cite{AEgIS_emul2}, although only 43 events were analysed in this study, and the tracking efficiency of the detector was limited to 80\%. 
In this paper, we present the results of a study of the multiplicities of charged annihilation products on different target materials, namely copper, silver and gold, using emulsion detectors at the Antiproton Decelerator (AD~\cite{AD}) at CERN.

\section{Experimental setup}

\begin{figure}
\centering
\includegraphics[width=1\textwidth]{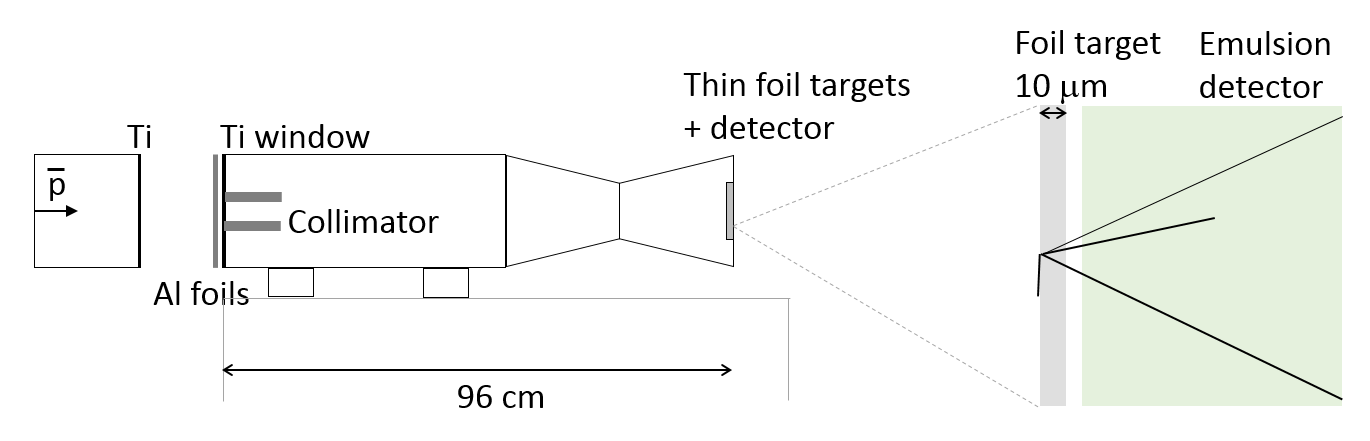}
\caption{Schematic setup of the experiment. An enlarged view of the target region is shown on the right.
} \label{fig:setup}
\end{figure}

Emulsion detectors were used to study the antiproton annihilation products generated in different materials. Before reaching the targets, the 5.3 MeV antiprotons from the AD ($3 \times 10^7$ $\overline{p}$/shot every 100 s) were slowed down using several different titanium and aluminium foils with variable thicknesses. Finally, the beam was collimated in a vacuum test chamber after crossing a titanium vacuum separation window with a thickness of 12~$\mu$m. 
The emulsion detector was situated at the downstream end of the vacuum chamber ($\sim$1~m in length), where it could be reached by a defocused beam of low-energy antiprotons ($\sim$100~keV). This distance from the degrading layer was necessary to reduce the background due to annihilations taking place at the moderator. 
A sketch of the experimental assembly is shown in Figure~\ref{fig:setup}. The emulsion detector was operated under ordinary vacuum conditions ($10^{-5}-10^{-6}$ mbar). The antiproton intensity measured by the detector was approximately 150/cm$^2$ per shot.

For this study emulsion detectors were produced at the Laboratory for High Energy Physics (LHEP) of the University of Bern by pouring the emulsion gel with a thickness of $\sim$100~$\mu$m, provided by Nagoya University (Japan), on a glass plate (for a review on the emulsion technology see ~\cite{ereditato}). Glycerin was added to so that the emulsion could operate in vacuum~\cite{AEgIS_emul2}. This emulsion features a very low background with approximately 1-2 thermally induced grains per 1000 $\mu$m$^{3}$~\cite{quplas}.

\begin{figure}
\begin{center}
\begin{minipage}{0.6\textwidth}
\includegraphics[width=1\textwidth]{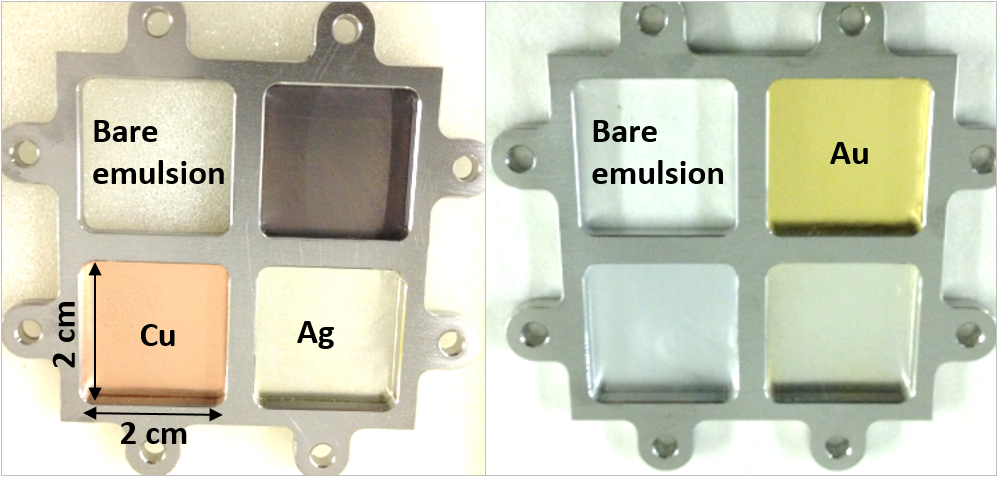}
\end{minipage}
\begin{minipage}{0.35\textwidth}
\includegraphics[width=1\textwidth]{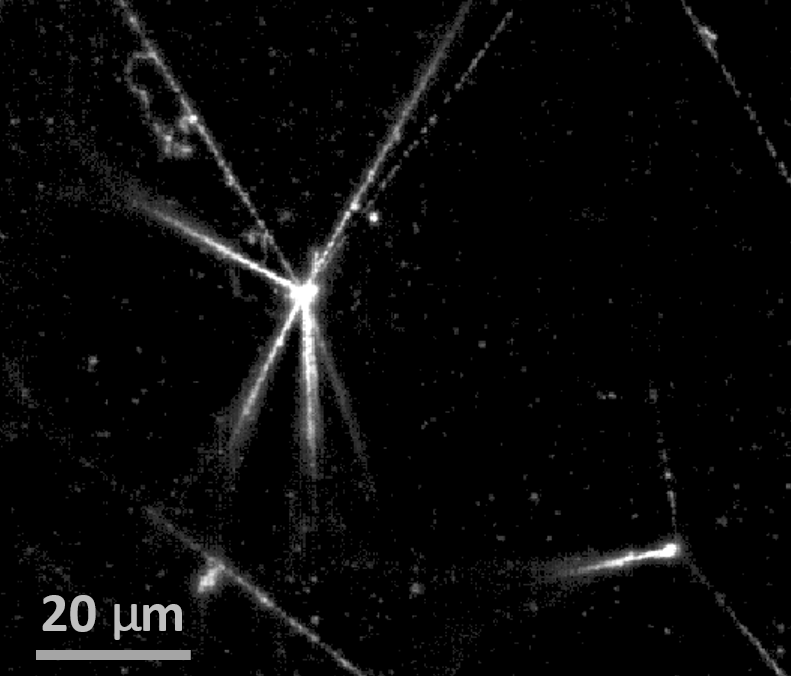}
\end{minipage}
\caption{Left and middle: Target arrangement in the two assemblies fixed to the emulsion films. Right: Antiproton annihilations on the bare emulsion surface.
} \label{fig:frame_and_bare_emul}
\end{center}
\end{figure}

Foils of copper, silver and gold, each having a thickness of 10~$\mu$m, were placed as targets at the end of the vacuum chamber, in front of the emulsion detectors. Figure~\ref{fig:frame_and_bare_emul} shows the targets ($2 \times 2$ cm$^{2}$ each) fixed to the emulsion film and an example of antiproton annihilating on the emulsion surface. At the antiproton energies obtained after degrading, all annihilations are expected to take place within a few $\mu$m of the target surface. During data taking, we collected approximately 1500 antiprotons per cm$^2$ in about 10 AD shots.

\section{Data analysis and results}

Data recorded by the emulsion detectors were scanned by an automatic optical microscope and then analysed by exploiting a recently developed fast 3D tracking algorithm~\cite{atmic}. The measured tracking efficiency of our detector was approximately 99\% for minimally ionizing particles over a wide angular range, as reported in~\cite{atmic}.

\begin{figure}
\centering
\includegraphics[width=0.7\textwidth]{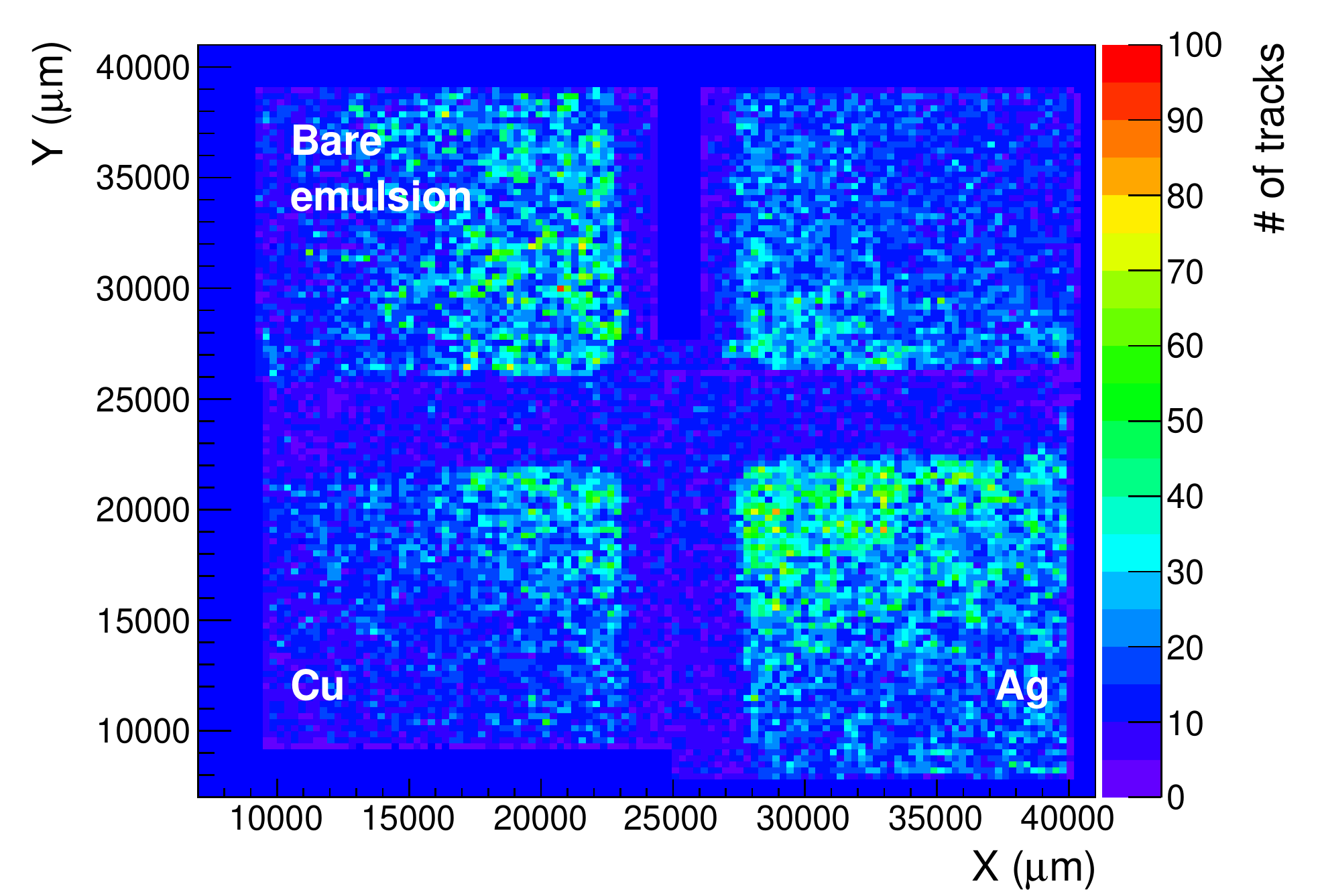}
\caption{
Profile of detected tracks ($xy$ positions of tracks at the emulsion layer) in one of the assemblies. The top right part was not included in our study. 
} \label{fig:trackdensity}
\end{figure}

Figure~\ref{fig:trackdensity} shows the profile of detected tracks ($xy$ positions of tracks at the emulsion layer) in one of the assemblies. 
Among the reconstructed tracks we only considered those that were longer than 30 $\mu$m to avoid considering tracks that were due to the background. 
An angular cut of 0.4$<$tan$\theta$$<$2.0 (22$^\circ$$<$$\theta$$<$63$^\circ$) was applied, where $\theta$ is the track angle with respect to the beam direction. 
To reconstruct a vertex, at least two three-dimensional tracks were required. The efficiency of vertex reconstruction was estimated by applying the criteria given above to the output of the FLUKA simulation. It was found to be 22\% for copper, 24\% for silver, and 18\% for gold.

\begin{figure}
\centering
\includegraphics[width=0.75\textwidth]{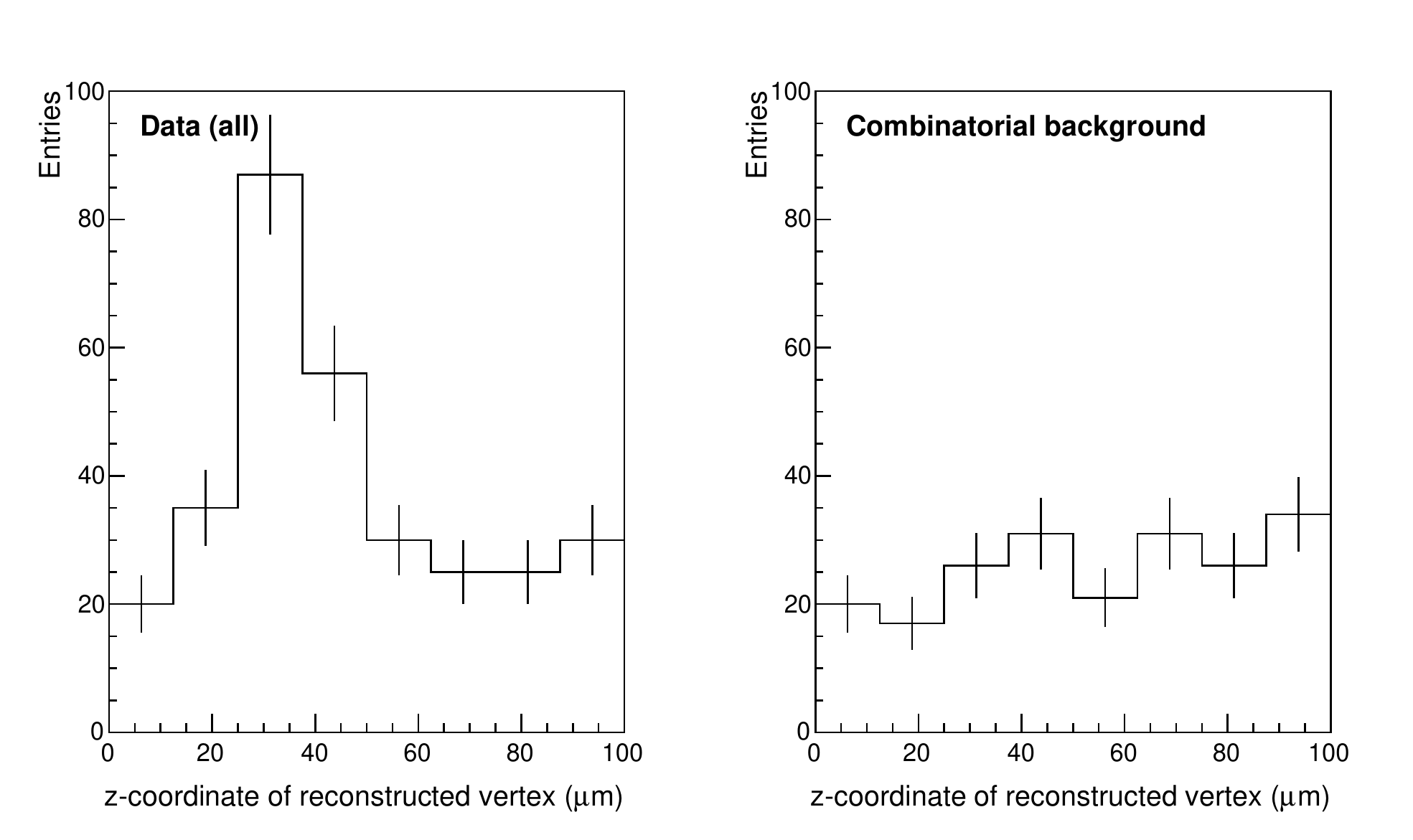}
\caption{Left: Distribution of the reconstructed vertex position perpendicular to the film surface (z-direction). Right: Estimated combinatorial background. 
} \label{fig:vz}
\end{figure}

\begin{figure}
\centering
\includegraphics[width=1.0\textwidth]{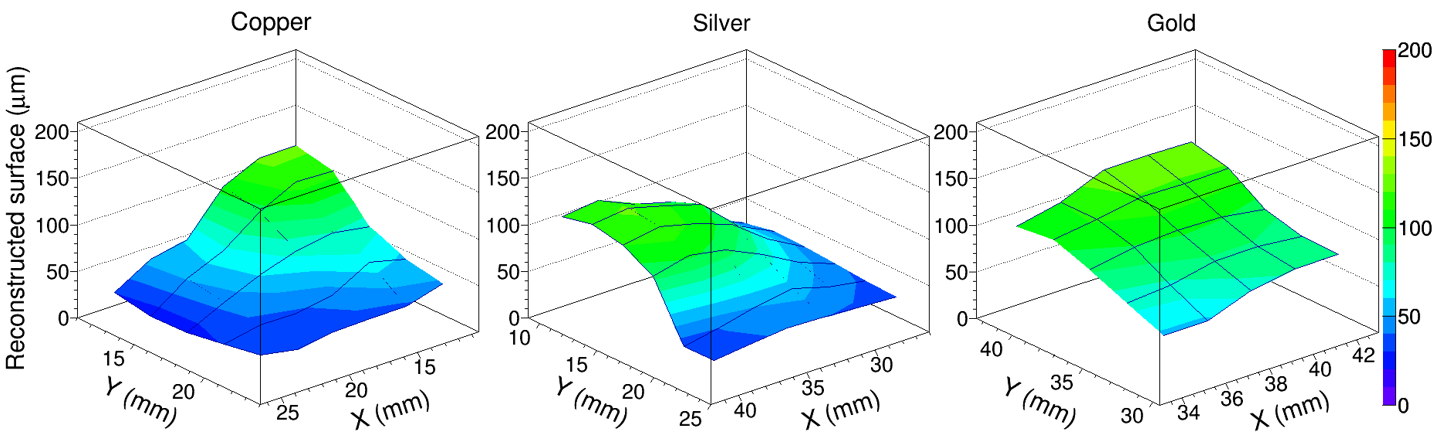}
\caption{Surface topography for copper, silver and gold targets obtained from the reconstructed vertices. The vertical scale refers to the distance from the emulsion film. 
} \label{fig:topo}
\end{figure}

Figure~\ref{fig:vz} (left) shows the distribution of the reconstructed vertex position perpendicular to the film surface ($z$-direction) in a subarea of the copper target. The peak in $z$ is a measure of the target foil position (the gap between the emulsion surface and the target foil). This measurement was performed by segmenting the analysed area into smaller areas since the target foils were neither flat nor in contact with the film surface. 
The surface topography obtained from the reconstructed vertices is shown in Figure~\ref{fig:topo} for copper, silver and gold targets. 
In our analysis, we only considered regions of the emulsion film surface with the $z$ value smaller than 100~$\mu$m because the vertex reconstruction efficiency was uniform within a few percent in this region. 
The analysed fiducial area was 1.68 cm$^2$ for the copper target, 1.96 cm$^2$ for silver and 0.80 cm$^2$ for gold. 
The fraction of signal vertices became dominant by requiring vertex reconstruction at the position of the target foil. 

The nearly flat distribution in Figure~\ref{fig:vz} (right) is due to combinatorial background in the vertex reconstruction. The main source of the background was due to accidental combinations of tracks coming from annihilations taking place upstream in the apparatus, which were not completely excluded by the angular cut due to the broad angular distribution. The fraction of tracks from signal vertices to all detected tracks was estimated using the vertex finding efficiency described above and found to be 9\% for copper, 9\% for silver, and 6\% for gold. 
The number of background vertices was estimated using all the detected tracks in the analysed area by subtracting the above signal track fraction, randomizing positions and slopes of the remaining tracks, reconstructing the vertices and counting the number of vertices that mimicked annihilations in the target.

\begin{figure}[htbp]
\centering
\includegraphics[width=0.65\textwidth]{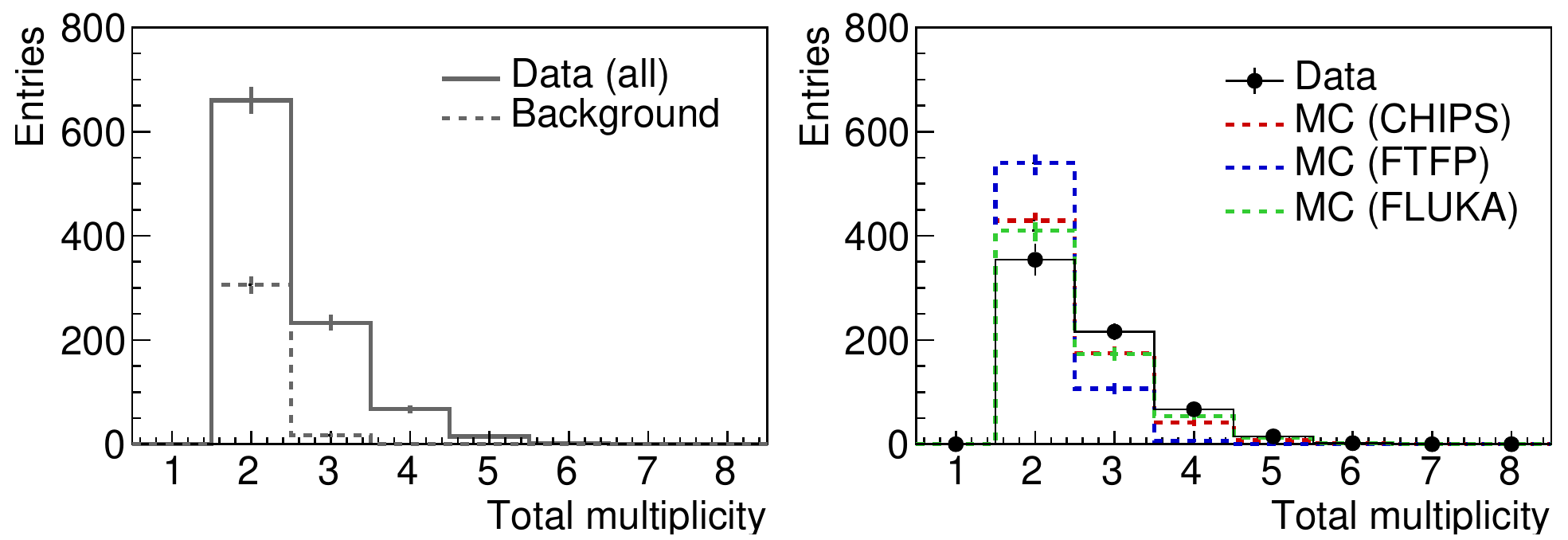}
\\
Copper
\\
\includegraphics[width=0.65\textwidth]{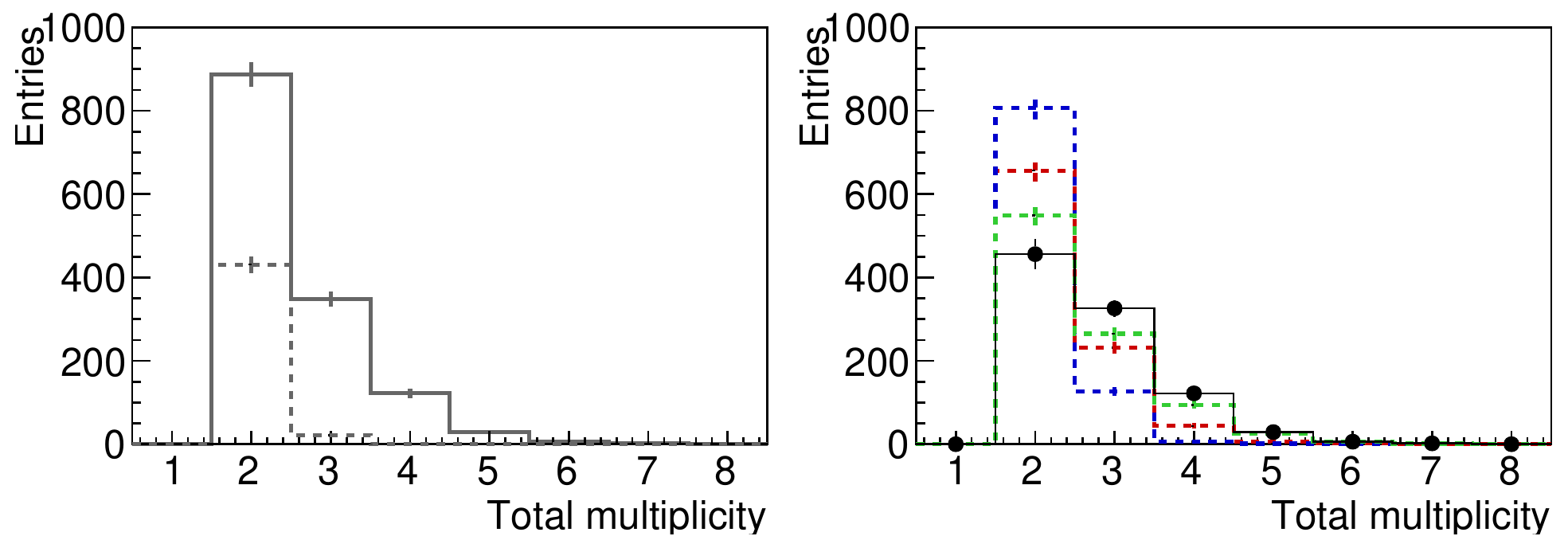}
\\
Silver
\\
\includegraphics[width=0.65\textwidth]{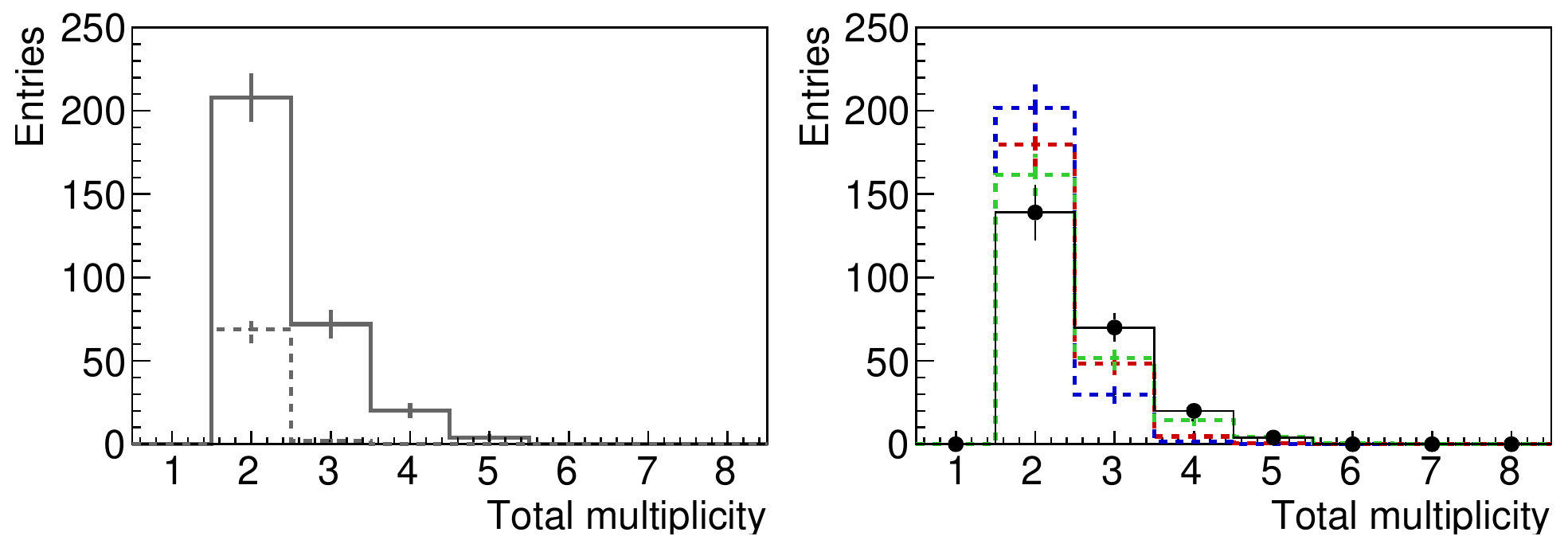}
\\
Gold
\caption{
Left: Background contributions to the total multiplicities for the different targets. Right: Multiplicity distributions after subtraction of the background. The colored histograms show the Monte Carlo predictions.
}\label{fig:ntot}
\end{figure}

The background estimated from measured data, which depends on the number of tracks in the event, is shown in the left panes of Figure~\ref{fig:ntot}, while the right panes show the multiplicity distributions after subtraction of the background, compared with the Monte Carlo predictions based on the CHIPS \cite{chips} and FTFP (FTFP\_BERT\_TRV) \cite{ftfp} models in the GEANT4 (4.9.5.p02) and FLUKA (2011.2c) \cite{fluka} frameworks. 
A total of 654 signal annihilation vertices were reconstructed for copper, 941 for silver and 233 for gold. 

\begin{figure}
\centering
\includegraphics[width=0.6\textwidth]{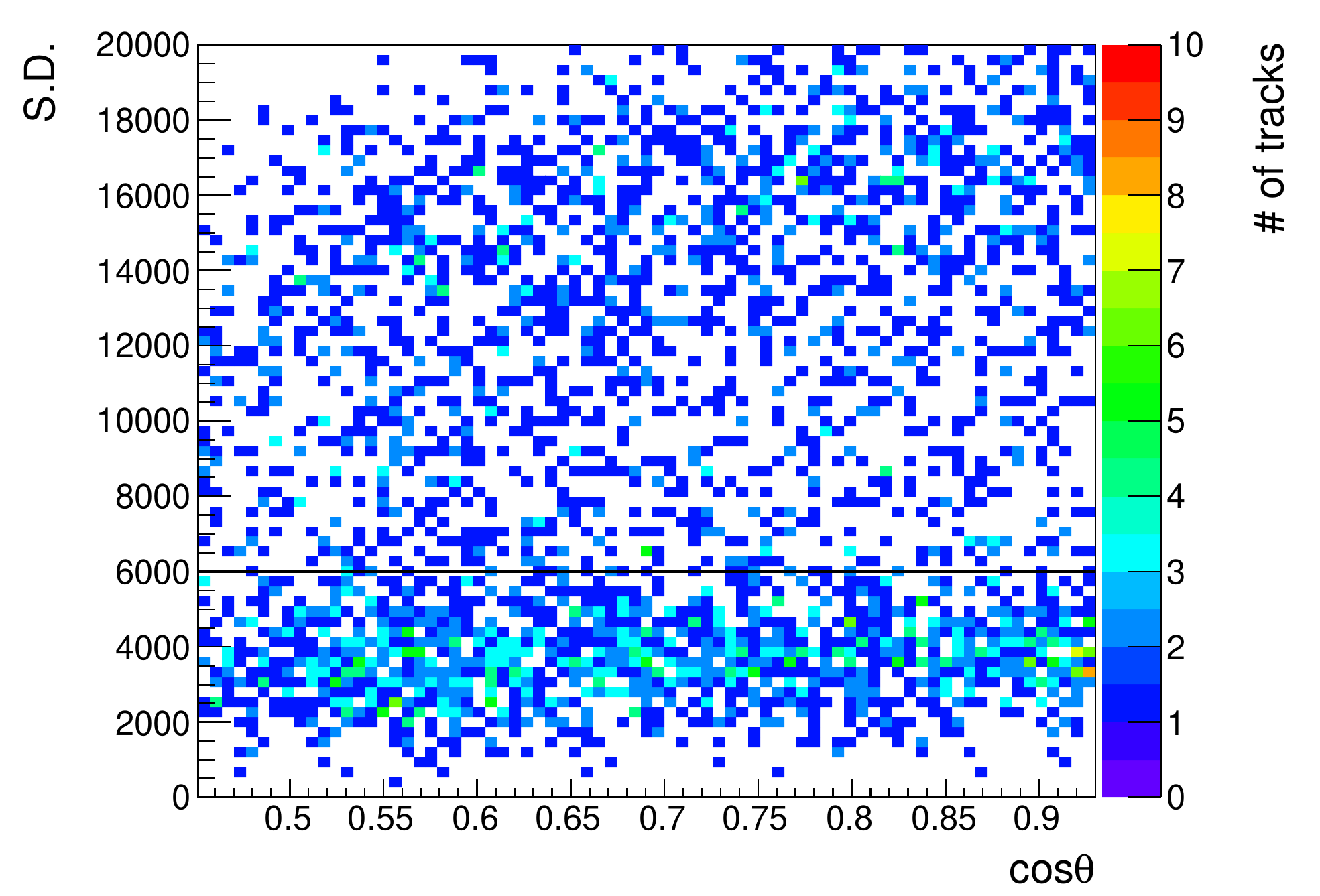}
\caption{Signal density (S.D.) distribution as a function of track angle with respect to the beam direction. Tracks below the black line are defined as being minimally ionizing.
} \label{fig:cos_sd}
\end{figure}

\begin{figure}[htbp]
\centering
\includegraphics[width=0.7\textwidth]{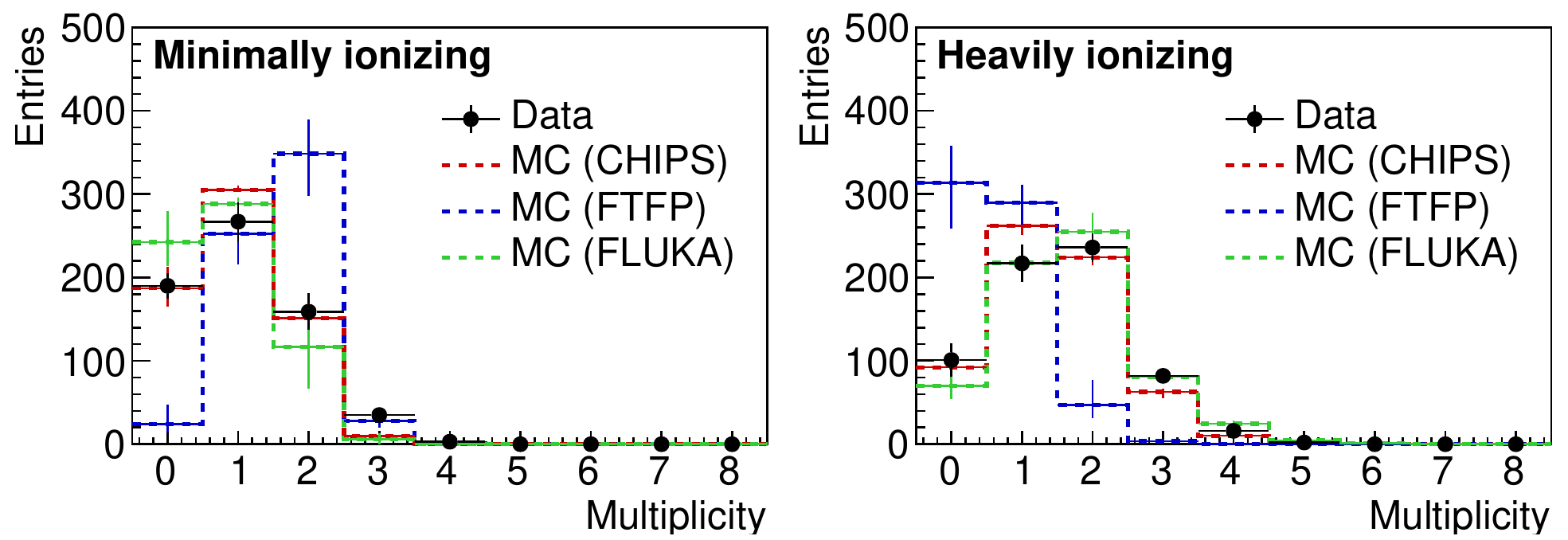}
\\
Copper
\\
\includegraphics[width=0.7\textwidth]{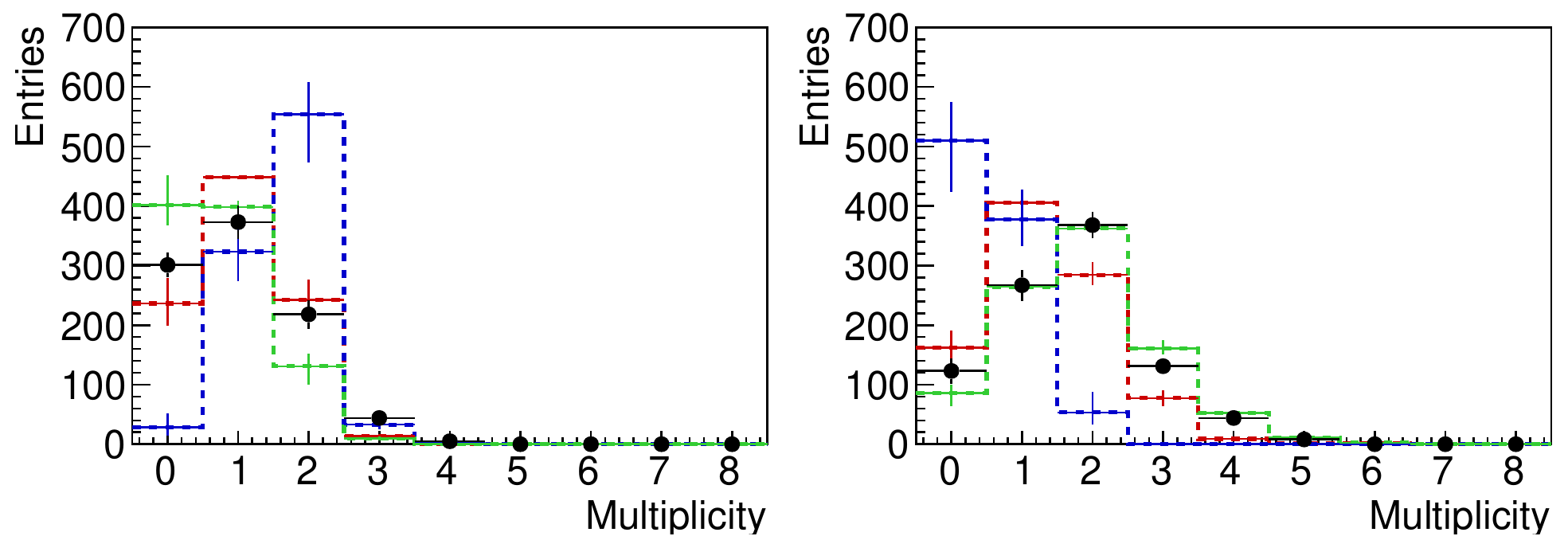}
\\
Silver
\\
\includegraphics[width=0.7\textwidth]{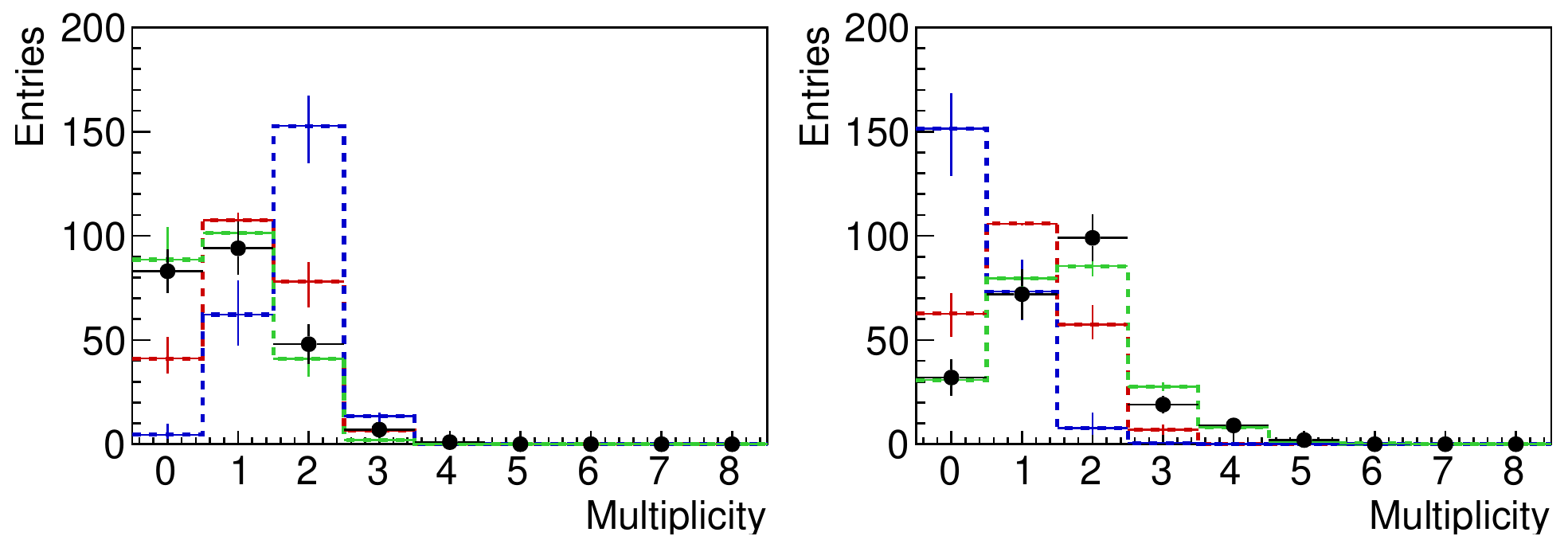}
\\
Gold
\caption{
Reconstructed multiplicity distributions for annihilations in copper, silver and gold foils for MIPs (left) and HIPs (right). The colored histograms show the Monte Carlo predictions by CHIPS, FTFP and FLUKA. The error bars of the histograms account for uncertainties in the $dE/dx$ classification.
}\label{fig:nsnh}
\end{figure}

\renewcommand{\arraystretch}{1.2}
\begin{table}[hbtb]
\footnotesize
\begin{center}
\begin{tabular}{|l|c|c|c|c|c|c|c|c|}
\hline
\multicolumn{1}{|l|}{} & \multicolumn{4}{c|}{Average multiplicity for MIPs} & \multicolumn{4}{c|}{Average multiplicity for HIPs} \\
\hline
                       & Data            & CHIPS                  & FTFP                   & FLUKA                  & Data            & CHIPS                  & FTFP                   & FLUKA \\
\hline
Copper                 & 1.07 $\pm$ 0.07 & 0.98$^{+0.07}_{-0.09}$ & 1.59$^{+0.09}_{-0.14}$ & 0.83$^{+0.08}_{-0.11}$ & 1.54 $\pm$ 0.07 & 1.46$^{+0.09}_{-0.07}$ & 0.60$^{+0.14}_{-0.10}$ & 1.68$^{+0.11}_{-0.08}$ \\
\hline
Silver                 & 1.02 $\pm$ 0.06 & 1.04$^{+0.08}_{-0.09}$ & 1.64$^{+0.09}_{-0.13}$ & 0.73$^{+0.07}_{-0.09}$ & 1.71 $\pm$ 0.07 & 1.33$^{+0.09}_{-0.08}$ & 0.51$^{+0.14}_{-0.09}$ & 1.87$^{+0.09}_{-0.07}$ \\
\hline
Gold                   & 0.92 $\pm$ 0.09 & 1.21$^{+0.10}_{-0.11}$ & 1.75$^{+0.09}_{-0.13}$ & 0.81$^{+0.07}_{-0.11}$ & 1.60 $\pm$ 0.09 & 1.04$^{+0.11}_{-0.09}$ & 0.39$^{+0.13}_{-0.09}$ & 1.60$^{+0.11}_{-0.06}$ \\
\hline
\end{tabular}
\caption{Measured average multiplicity and Monte Carlo predictions by CHIPS, FTFP (GEANT4) and FLUKA. Statistical errors and uncertainties in the background estimation are combined and reported for each set of data. The errors in the predictions account for uncertainties in $dE/dx$ classification.
}
\label{tb:ave}
\end{center}
\end{table}
\renewcommand{\arraystretch}{1}

We were also able to discriminate between heavily ionizing particles (HIPs) such as protons and nuclear fragments and minimally ionizing particles (MIPs), namely pions. 
Continuous dense tracks correspond to HIPs, while faint tracks are produced by MIPs, since the aligned grains of these last tracks are separated. 
The local energy deposition ($dE/dx$) of each track can then be assessed in terms of signal density (S.D.) along the reconstructed tracks, using 
\[
S.D. = \sum_{x,y,z\in C}S_{xyz}/L.
\]
Here, $x$, $y$ and $z$ are the coordinates of voxels in the 3D image data. $C$ is a group of voxels in a cylinder along the track, and $S_{xyz}$ is an 8-bit grey-scale signal of the voxel. $L$ is the length of a reconstructed track in the 3D image data. The S.D. is proportional to the $dE/dx$ of the particle and does not depend on the angle. However, there is a saturation effect for higher values of $dE/dx$.
Figure~\ref{fig:cos_sd} shows that the S.D. distribution of tracks revealed a peak at 3000 for MIPs. 
As the simulated $dE/dx$ distribution of MIPs peaked at 1.2 MeV$\cdot$g$\cdot$cm$^{-2}$, we define particles with $dE/dx$ smaller than 2.4 MeV$\cdot$g$\cdot$cm$^{-2}$, corresponding to an S.D. below 6000 $\mu$m$^{-1}$, as being MIPs. The complementary particles are defined as HIPs. 

Figure~\ref{fig:nsnh} shows the track multiplicity distributions for MIPs and HIPs. The errors on the data are statistical. 
The histograms represent the Monte Carlo predictions by CHIPS, FTFP and FLUKA. The error bars of the Monte Carlo predictions account for uncertainties in the $dE/dx$ classification. These uncertainties were estimated using simulations, which smeared the threshold for assigning tracks to either class of ionizing particles by 20\% and checked for effects on the multiplicity distributions for MIPs and HIPs. 
The statistical errors for the simulations were 0.01-0.02, which are significantly smaller than the errors reported above. 
The average measured multiplicities are summarized in Table~\ref{tb:ave}.
Both CHIPS and FLUKA are in good agreement for copper, particularly in the case of MIPs. Neither CHIPS nor FTFP accurately describe particle multiplicities for annihilations on silver and gold nuclei, while FLUKA more closely reproduce the data than the other models.

\begin{figure}[htbp]
\centering
\includegraphics[width=0.65\textwidth]{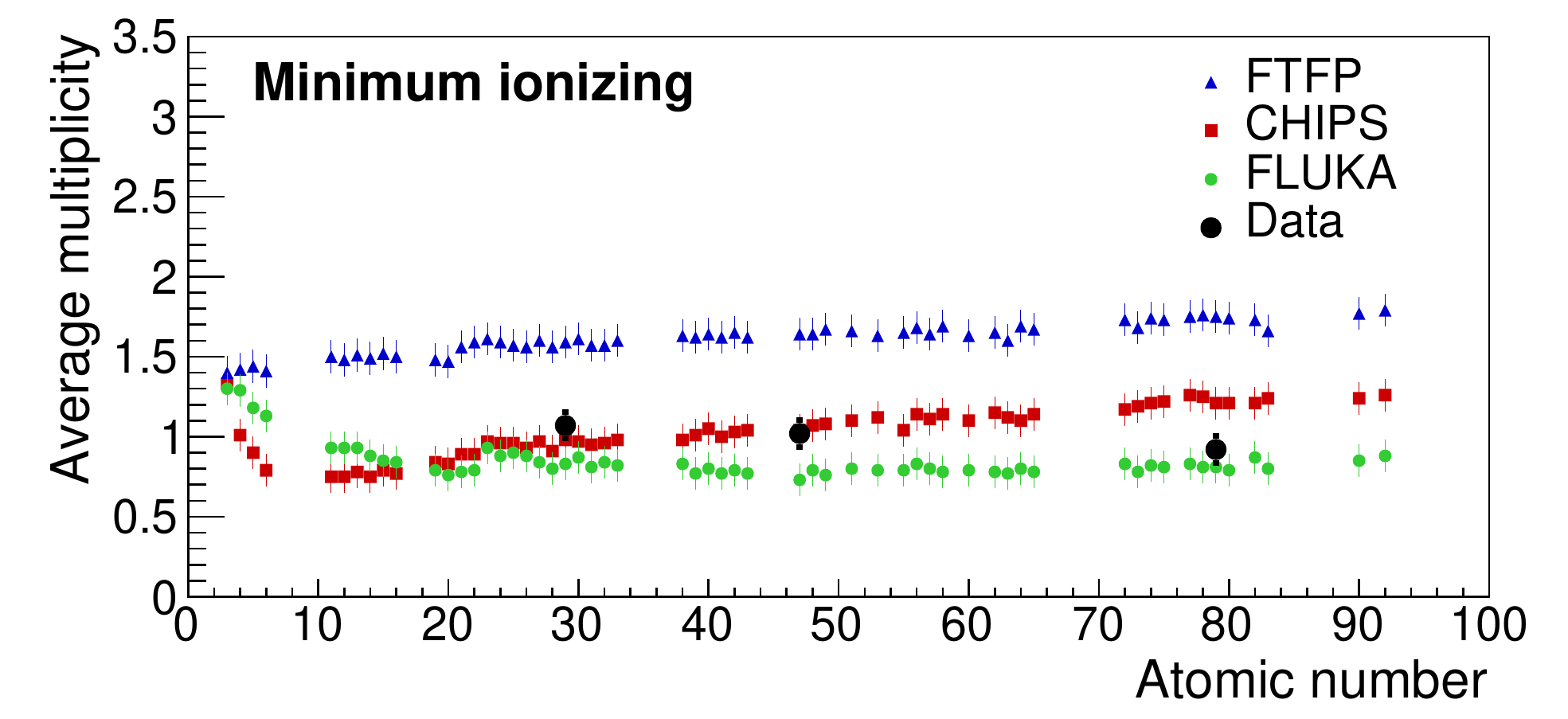}
\\
\includegraphics[width=0.65\textwidth]{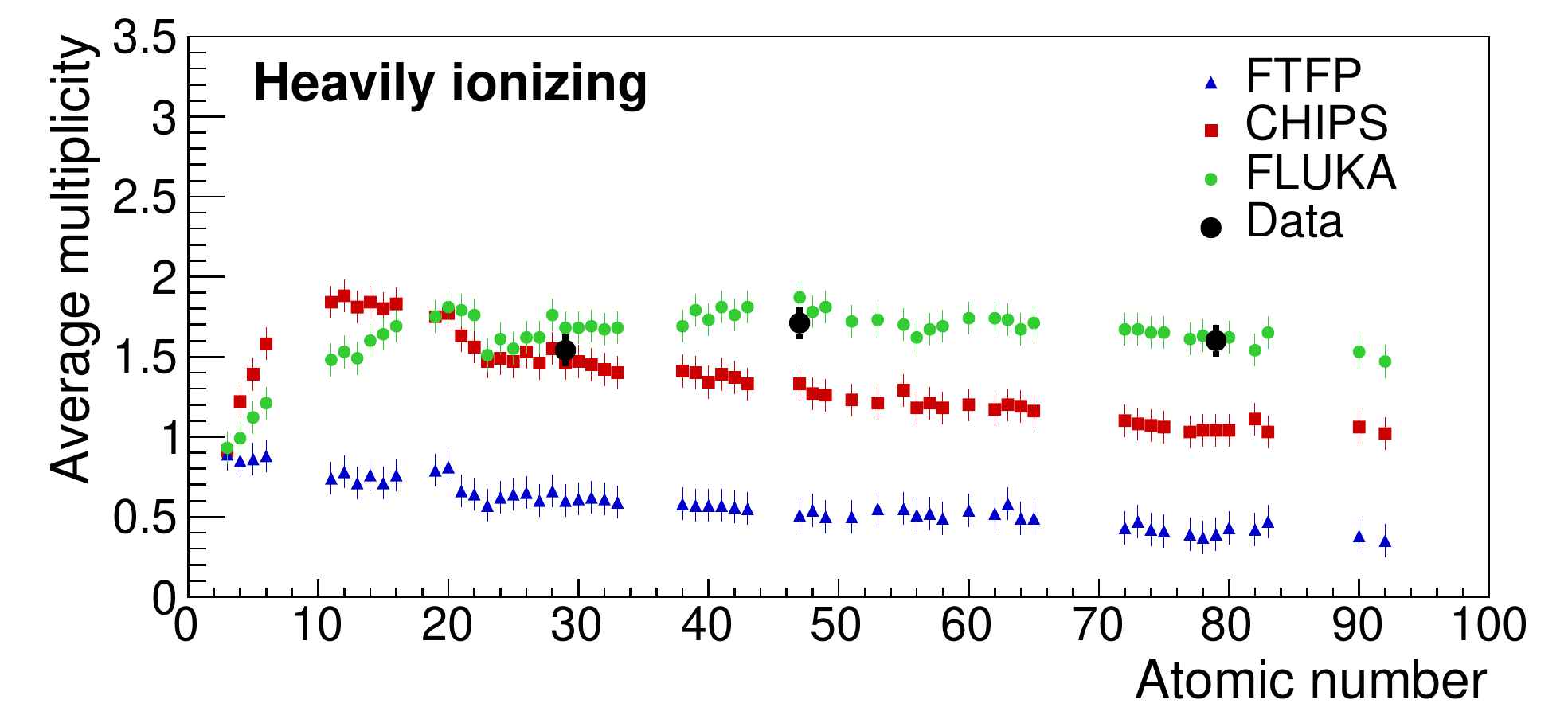}
\caption{Particle multiplicity from antiproton annihilations as a function of atomic number for MIPs (top) and HIPs (bottom).
} \label{fig:A_dep}
\end{figure}

The mean values of particle multiplicities measured for the three target materials are shown in Figure~\ref{fig:A_dep} as a function of atomic number along with the simulation outcome. Results obtained for MIPs with the FTFP model do not agree with our experimental data for any material, while those obtained with both CHIPS and FLUKA are in fair agreement as far as copper is concerned, although only FLUKA reproduces the higher atomic number behavior. Good agreement with CHIPS was also found for annihilation on bare emulsions and for aluminium~\cite{AEgIS_emul2}. Multiplicities related to HIPs are well described by the FLUKA simulation, while the CHIPS and FTFP models clearly underestimate the number of particles produced by antiproton annihilation.

\section{Conclusions}
The goal of the study presented in this paper was to measure the products of low-energy antiproton annihilation on different materials, utilizing a secondary beam line of CERN's Antiproton Decelerator in the AEgIS experimental area. The characteristics (e.g. hadronization and fragmentation multiplicities) of low-energy antiprotons annihilating on nuclei are not well known, and experimental data are needed to validate models used by simulation packages such as GEANT4 and FLUKA. 
We exposed several thin targets (Cu, Ag and Au) to the antiproton beam and measured fragment tracks using emulsion detectors with a vertex position resolution at the level of a few micrometers, which allowed the separation between minimally and highly ionizing particles. The fragment multiplicities we measured were not well reproduced by the different models used in Monte Carlo simulation with the exception of FLUKA, which is in good agreement with the particle multiplicities for both minimally and heavily ionizing particles. Future measurements with more materials are needed to gain a better understanding of antinucleon annihilations also on low-$Z$ materials, and to obtain a full description in terms of particle types, multiplicities, as well as energy, for a more complete benchmarking of Monte Carlo simulations.\\

\acknowledgments
This work was supported by the Swiss National Science Foundation Ambizione grant PZ00P2\_154833; Istituto Nazionale di Fisica Nucleare; a Deutsche Forschungsgemeinschaft research grant; an excellence initiative of Heidelberg University; European Research Council under the European Unions Seventh Framework Program FP7/2007-2013 (Grants No. 291242 and No. 277762); Austrian Ministry for Science, Research, and Economy; Research Council of Norway; Bergen Research Foundation; John Templeton Foundation; Ministry of Education and Science of the Russian Federation and Russian Academy of Sciences; and the European Social Fund within the framework of realizing the project, in support of intersectoral mobility and quality enhancement of research teams at Czech Technical University in Prague (Grant No. CZ.1.07/2.3.00/30.0034). The authors would like to acknowledge the contributions by the mechanical workshop at LHEP.


\begin{thebibliography}{99}

\bibitem{aegisprop} M. Doser et al., \emph{Exploring the WEP with a pulsed cold beam of antihydrogen}, \emph{Class. Quant. Grav.} \textbf{29} (2012) 184009.

\bibitem{AEgIS_emul1} C. Amsler et al., \emph{A new application of emulsions to measure the gravitational force on antihydrogen}, \jinst{8}{2013}{P02015}. 

\bibitem{AEgIS_emul2} S. Aghion et al. (AEgIS Collaboration), \emph{Prospects for measuring the gravitational free-fall of antihydrogen with emulsion detectors}, \jinst{8}{2013}{P08013}. 

\bibitem{nature} S. Aghion et al. (AEgIS Collaboration), \emph{A moir\'e deflectometer for antimatter}, \emph{Nature Communications} \textbf{5} (2014) 4538. 

\bibitem{quplas} S. Aghion et al., \emph{Emulsion films for low energy antimatter detection}, \jinst{11}{2016}{P06017}.

\bibitem{ereditato} A. Ereditato, \emph{The Study of Neutrino Oscillations with Emulsion Detectors}, \emph{Adv. High Energy Phys.} \textbf{2013} (2013) 382172. 

\bibitem{silicon} S. Aghion et al. (AEgIS Collaboration), \emph{Detection of low energy antiproton annihilations in a segmented silicon detector}, \jinst{9}{2014}{P06020}. 

\bibitem{ref1994} G. Bendiscioli and D. Kharzeev, \emph{Antinucleon-nucleon and antinucleon-nucleus interaction. A review of experimental data}, \emph{Riv. Nuovo Cim.} \textbf{17N6} (1994) 1-142.

\bibitem{ref1995} D. Polster et al., \emph{Light particle emission induced by stopped anti-protons in nuclei: Energy dissipation and neutron to proton ratio}, \emph{Phys. Rev. C} \textbf{51} (1995) 1167-1180.

\bibitem{AD} https://home.cern/about/accelerators/antiproton-decelerator.

\bibitem{atmic} A. Ariga and T. Ariga, \emph{Fast 4$\pi$ track reconstruction in nuclear emulsion detectors based on GPU technology}, \jinst{9}{2014}{P04002}. 

\bibitem{chips} P.V. Degtyarenko, M. Kosov and H. Wellisch, 
\emph{Chiral invariant phase space event generator. I: Nucleon antinucleon annihilation at rest}, \emph{Eur. Phys. J. A} \textbf{8} (2000) 217; 
\emph{Chiral invariant phase space event generator. II: Nuclear pion capture at rest and photonuclear reactions below the Delta(3,3) resonance}, \emph{Eur. Phys. J. A} \textbf{9} (2000) 411; 
\emph{Chiral invariant phase space event generator. III: Modeling of real and virtual photon interactions with nuclei below pion production threshold}, \emph{Eur. Phys. J. A} \textbf{9} (2000) 421; 
\emph{M. Kossov, Simulation of antiproton-nuclear annihilation at rest}, IEEE Trans. Nucl. Sci. 52 (2005) 2832.

\bibitem{ftfp} 
B. Andersson, G. Gustafson and B. Nilsson-Almqvist, \emph{A model for low-pT hadronic reactions with generalizations to hadron-nucleus and nucleus-nucleus collisions}, \emph{Nucl. Phys. B} \textbf{281} (1987) 289; 
B. Nilsson-Almqvist and E. Stenlund, \emph{Interactions between hadrons and nuclei: The Lund Monte Carlo - FRITIOF version 1.6 -}, \emph{Comput. Phys. Commun.} \textbf{43} (1987) 387. 

\bibitem{fluka}
A. Ferrari, P.R. Sala, A. Fasso, and J. Ranft, \emph{FLUKA: a multi-particle transport code}, CERN 2005-10 (2005), INFN/TC\_05/11, SLAC-R-773; 
T.T. Bohlen, F. Cerutti, M.P.W. Chin, A. Fasso, A. Ferrari, P.G. Ortega, A. Mairani, P.R. Sala, G. Smirnov, and V. Vlachoudis, \emph{The FLUKA Code: Developments and Challenges for High Energy and Medical Applications}, Nuclear Data Sheets \textbf{120} (2014) 211-214.

\end{thebibliography}
\end{document}